\documentclass{emulateapj}
\usepackage{apjfonts}

\slugcomment{ApJL in press}

\shorttitle{2D PIC Simulation of a Perpendicular Shock}
\shortauthors{Umeda, Yamao \& Yamazaki}

\begin{document}

\title{
Two-dimensional full particle simulation of a
perpendicular collisionless shock 
with a shock-rest-frame model
}

\author{
Takayuki~Umeda\altaffilmark{1}, 
Masahiro~Yamao\altaffilmark{1}, 
and Ryo~Yamazaki\altaffilmark{2} 
}
\altaffiltext{1}{
Solar-Terrestrial Environment Laboratory, Nagoya University, 
Nagoya, 464-8601, Japan; umeda@stelab.nagoya-u.ac.jp}
\altaffiltext{2}{Department of Physical Science, Hiroshima University,
Higashi-Hiroshima 739-8526, Japan; ryo@theo.phys.sci.hiroshima-u.ac.jp}
%


\begin{abstract}
A two-dimensional (2D) shock-rest-frame model 
for particle simulations is developed. 
Then full kinetic dynamics of a perpendicular collisionless shock 
is examined 
by means of a 2D full particle simulation. 
We found that in the 2D simulation 
there are fewer  nonthermal electrons 
due to surfing acceleration 
which was seen in the previous 1D simulations 
of a high Mach number perpendicular shock 
in a low-beta and weakly magnetized plasma. 
This is because 
the particle motion along the ambient magnetic field 
disturbs the formation of coherent electrostatic solitary structures 
which is necessary for electron surfing acceleration. 
\end{abstract}



\section{Introduction}

Observations of shell-type supernova remnants (SNRs)
provide us evidence for electron acceleration at SNR
shocks, and in some cases the maximum energy of the electrons
exceeds $\sim$10~TeV \citep{koyama1995,bamba2003,bamba2005}. 
The most plausible mechanism of the acceleration is the diffusive
shock acceleration (DSA), which explains the broadband power-law spectrum
with an index around 2 \citep{drury1983,blandford1987}.
Before the DSA phase in which electrons cross the shock front many
times, they have to be pre-accelerated by an unknown injection mechanism.
This ``injection problem'' is still unresolved.

One of the possible injection mechanisms is the shock surfing
acceleration \citep{Shimada_2000,Hoshino_2002}. 
\citet{Shimada_2000} performed 
a one-dimensional (1D) electromagnetic full particle simulation
and found 
the formation of large-amplitude electrostatic solitary structures 
during the cyclic reformation of 
a high Mach number perpendicular shock in a low-beta and 
weakly magnetized plasma. 
The coherent solitary structures 
trap electrons in their electrostatic potential well, 
resulting in significant surfing acceleration of electrons.
The electron surfing has been regarded as 
an efficient direct acceleration mechanism at SNR shocks 
and studied by many authors 
\citep[e.g.,][]{McClements_2001,Hoshino_2002,
Schmitz_2002a,Schmitz_2002b,Amano_2007}.
However, these previous works are based on 1D simulations, 
in which configuration of shock magnetic fields 
cannot be modified. 
Since charged particles can easily drift along magnetic fields, 
it is important to take into account 
the modification of shock magnetic fields 
and the free motion of particles along shock magnetic fields. 
Hence, multi-dimensional simulations are necessary.

In the present study we perform a 2D 
full particle simulation of a high Mach number and low-beta 
perpendicular collisionless shock. 
The preliminary result shows that the electron surfing acceleration 
is suppressed in a 2D system.

\section{Full Particle Simulation}

Generally speaking, it is difficult to perform multi-dimensional 
simulations of collisionless shocks 
even with present-day supercomputers.
In the previous simulations of high Mach number shocks 
\citep[e.g.,][]{Shimada_2000,Schmitz_2002a,Schmitz_2002b,
Lee_2004,Amano_2007}, 
a collisionless shock is excited with 
the ``injection method'' 
(also called the reflection, wall, or piston method), 
in which a plasma is injected from one side of
the simulation domain and is reflected back when it reaches
the other side. 
A collisionless shock is excited by interaction 
between injected and reflected plasmas. 
Thus, the simulation domain 
is taken in the downstream rest frame, 
and the excited shock wave propagates upstream. 
It is necessary to take a very long simulation
domain in the propagation direction of the shock wave
in order to study long-time evolution of moving shock waves.

Recently, 1D full particle simulations have been 
performed in an almost shock rest frame 
\citep{Muschietti_2006,Umeda_2006}. 
In these simulations, 
a collisionless shock is excited with 
the ``relaxation method,'' 
which was first used in hybrid code simulations 
\citep{Leroy_1981,Leroy_1982}. 
\citet{Muschietti_2006} used the Darwin approximation
in which the displacement current 
due to the transverse electric field is neglected, 
while \citet{Umeda_2006} solved the full set of 
Maxwell's equations to 
take into account all the plasma waves. 
Although the shock-rest-frame model is useful to perform 
simulations of shocks with less computer resources, 
there have been few multi-dimensional shock-rest-frame models. 
Here, 
we successfully extended the previous 1D shock-rest-frame model 
\citep{Umeda_2006} to two dimensions.

We use a 2D electromagnetic particle code 
\citep{Umeda_PhD}, in which 
the full set of Maxwell's equations and 
the relativistic equation of motion for individual electrons and ions 
are solved in a self-consistent manner. 
The continuity equation for charge is also 
solved to compute the exact current density 
given by the motion of charged particles \citep{Umeda_2003}. 

The initial state consists of two uniform regions 
separated by a discontinuity. 
In the upstream region that is taken in the left-hand side 
of the simulation domain, 
electrons and ions are distributed uniformly in space and 
are given random velocities $(v_x,v_y,v_z)$ to approximate 
shifted Maxwellian momentum distributions 
with the drift velocity $u_{x1}$, 
number density $n_{1} \equiv \epsilon_0 m_e \omega_{pe1}^2 / e^2$, 
and isotropic temperatures $T_{e1} \equiv m_e v_{te1}^2$ and 
$T_{i1} \equiv m_i v_{ti1}^2$, 
where $m$, $e$, $\omega_{p}$, and $v_{t}$ are 
the mass, charge, plasma frequency, and 
thermal velocity, respectively. 
Subscripts ``1'' and ``2'' denote 
``upstream'' and ``downstream,'' respectively.
The upstream magnetic field $B_{y01} \equiv -m_e \omega_{ce1}/e$ 
is also assumed to be uniform, where $\omega_{c}$ 
is the cyclotron frequency (with sign included). 
The downstream region taken in the right-hand side 
of the simulation domain is prepared similarly with 
the drift velocity $u_{x2}$, density $n_{2}$, 
isotropic temperatures $T_{e2}$ and $T_{i2}$, 
and magnetic field $B_{y02}$.

We take 
the simulation domain in the $x$-$y$ plane 
and assume a perpendicular shock (i.e., $B_{x0}=0$). 
Since the ambient magnetic field is taken in the $y$-direction, 
free motion of particles along the ambient magnetic field 
is taken into account. 
As a motional electric field, a uniform external electric field 
$E_{z0} =-u_{x1}B_{y01} =-u_{x2}B_{y02}$ 
is applied in both upstream and downstream regions, 
so that both electrons and ions drift in the $x$-direction. 
At the left boundary of the simulation domain in the $x$-direction,
we inject plasmas with the same quantities 
as those in the upstream region, 
while plasmas with the same quantities as those 
in the downstream region are also injected from the right boundary 
in the $x$-direction.
We adopted absorbing boundaries 
to suppress nonphysical reflection of electromagnetic waves at 
both ends of the simulation domain in the $x$-direction \citep{Umeda_2001}, 
while the periodic boundaries are imposed 
in the $y$-direction. 

The initial downstream quantities 
are given by solving 
the shock jump conditions for 
a magnetized two-fluid isotropic plasma
consisting of electrons and ions 
\citep[e.g.,][]{Hudson_1970}. 
In order to determine a unique initial downstream state, 
we need given upstream quantities 
$u_{x1}$, $\omega_{pe1}$, $\omega_{ce1}$, $v_{te1}$, and 
$T_{i1}/T_{e1}$ 
and another parameter. 
We assume a low-beta and weakly magnetized plasma 
such that $\beta_{e1}=\beta_{i1}=0.125$ and 
$\omega_{ce1}/\omega_{pe1}=-0.05$ in the upstream region. 
We also use a reduced ion-to-electron mass ratio $m_i/m_e = 25$ 
for computational efficiency. 
The light speed $c/v_{te1}=80.0$ and 
the bulk flow velocity of the upstream plasma $u_{x1}/v_{te1}=8.0$ 
are also assumed.
Then, the Alfv\'{e}n Mach number is calculated as
$M_A = (u_{x1}/c)|\omega_{pe1}/\omega_{ce1}|\sqrt{m_i/m_e}=10.0$. 
The ion-to-electron temperature ratio 
in the upstream region is given as $T_{i1}/T_{e1}=1.0$. 
Note that these parameters 
are almost the same as the previous full particle simulations 
of a perpendicular shock 
\citep{Shimada_2000,Hoshino_2002,Schmitz_2002a,Schmitz_2002b,Lee_2004}. 
In this study, the downstream ion-to-electron temperature ratio 
$T_{i2}/T_{e2} = 8.0$ is also assumed as 
an initial parameter 
to obtain the unique downstream quantities 
$\omega_{pe2}/\omega_{pe1} = 1.953$, 
$\omega_{ce2}/\omega_{pe1} = 0.191$, 
$u_{x2}/v_{te1} = 2.097$, 
and $v_{te2}/v_{te1} = 5.715$. 

We used $N_x \times N_y = 4096 \times 64$ 
cells for the upstream region and 
$N_x \times N_y = 2048 \times 64$ 
cells for the downstream region.
The grid spacing and time step of the present simulation are 
$\Delta x/\lambda_{e1} = 1.0$ and $\omega_{pe1}\Delta t=0.005$.
We used 16 pairs of electrons and ions per cell in the upstream region 
and 64 pairs of electrons and ions per cell in the downstream region 
 at the initial state. 
One might think that the downstream system size is too short 
because the downstream does have an affect on the transition region. 
However, 
we have confirmed 
by changing the downstream system size 
that there is not any critical change 
in the simulation result 
when the downstream system size is longer than 
the typical ion gyroradius 
$u_{x1}/\omega_{ci2}$ \citep{Umeda_2006}. 
This implies that 
the results are most appropriate to the transition region, 
and that the transition region, is approximately modeled 
by the present shock-rest-frame model 
despite the lack of an extensive downstream region. 
Thus, the processes seen in the present simulation should be robust.

\section{Result}

\begin{figure}[t]
\includegraphics[width=0.5\textwidth]{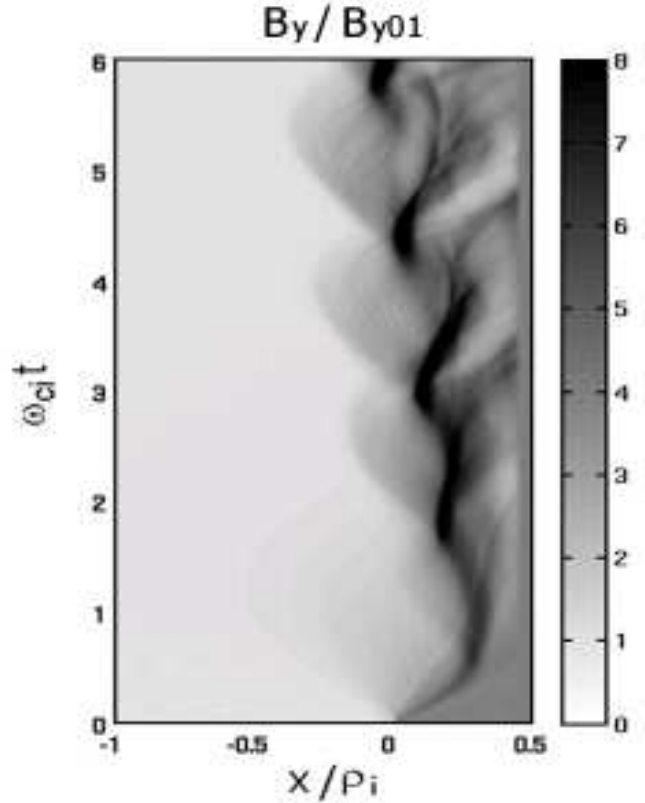}
\caption{
Transverse magnetic field $B_y$ 
as a function of position $x$ and time $t$ 
on the $x$-axis ($y=0$). 
The position and time are normalized by 
$\rho_i \equiv u_{x1}/\omega_{ci1}$ and $1/\omega_{ci1}$, 
respectively. 
The magnitude is 
normalized by the initial upstream magnetic field $B_{y01}$. 
}
\end{figure}

Figure 1 shows the transverse magnetic field $B_y$ 
as a function of position $x$ and time $t$ 
on the $x$-axis ($y=0$). 
In the present shock-rest-frame model, 
a shock wave is excited 
by the relaxation of the two plasmas with different quantities. 
Figure 1 shows that the shock front 
appears and disappears at a timescale of the downstream 
ion gyro-period, which corresponds to the 
cyclic reformation of high Mach number perpendicular shocks. 
Since the initial state is given by the shock jump conditions 
for a two-fluid plasma consisting of electrons and ions, 
the excited shock is ``almost'' at rest in the simulation domain. 
The shock front shifts upstream at a very slow velocity of 
$\sim 0.06u_{x1}$, 
because dynamics of reflected ions is not taken into account 
in the two-fluid approximation. 
We have also performed a 1D simulation 
with the same parameter as the 2D simulation 
and confirmed that the period of cyclic reformation 
in both simulations is the same.


\begin{figure}[t]
\includegraphics[width=0.5\textwidth]{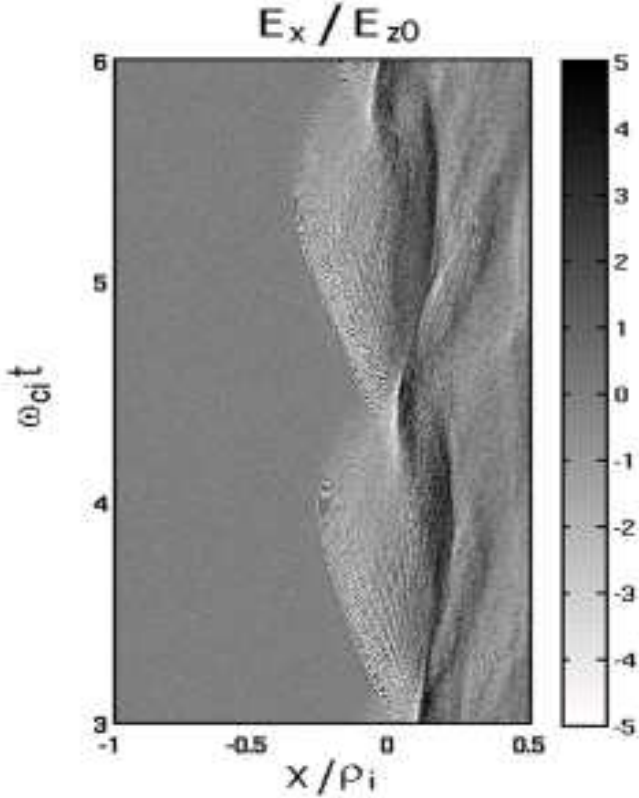}
\caption{
Shock-normal electric field $E_x$ 
as a function of position $x$ and time $t$ 
on the $x$-axis ($y=0$). 
The position and time are normalized by 
$\rho_i \equiv u_{x1}/\omega_{ci1}$ and $1/\omega_{ci1}$, 
respectively. 
The magnitude is 
normalized by the motional electric field $E_{z0}$. 
}
\end{figure}

Figure 2 shows the shock-normal electric field $E_x$ 
as a function of position $x$ and time $t$ 
on the $x$-axis ($y=0$). 
There is a good correlation between Figures 2 and 1, 
implying that the behavior of a shock-normal electric field 
is controlled by the shock reformation process. 
Figure 2 shows a bipolar signature of shock-normal electric field 
turning from negative to positive 
in the spatial scale of ion gyroradius 
from $x/\rho_i = -0.2$ to 0.2.  
The ion-scale electric field corresponds to 
the shock potential which reflects ions upstream. 
In the shock foot region 
there exist sinusoidal wave structures, 
which are excited by the current-driven instability 
due to reflected ions. 
These features are also seen in the 1D simulation 
(not shown here). 
However, 
there are two differences between the 
1D and 2D results. 
The first is that 
the amplitude of electric field 
in the 2D simulation is much smaller than 
that in the 1D simulation. 
The second is that 
the 1D simulation shows  
more complex wave structures with several large-amplitude pulses. 
The typical amplitude of coherent solitary structures 
in the previous 1D simulations 
was $E_x/E_{z0} \sim 20$ and the maximum amplitude 
exceeded $E_x/E_{z0} \sim 40$ \citep{Shimada_2000,Hoshino_2002}. 
In the present 2D simulation, 
on the other hand, 
the typical amplitude of a shock-normal electric field 
is $E_x/E_{z0} \sim 5$ 
and the maximum amplitude is $E_x/E_{z0} \sim 10$. 
This result suggests that 
the saturation level of the current-driven instability 
in the present 2D simulation is much lower than 
that in the previous 1D simulation.


\begin{figure}[t]
\includegraphics[width=0.5\textwidth]{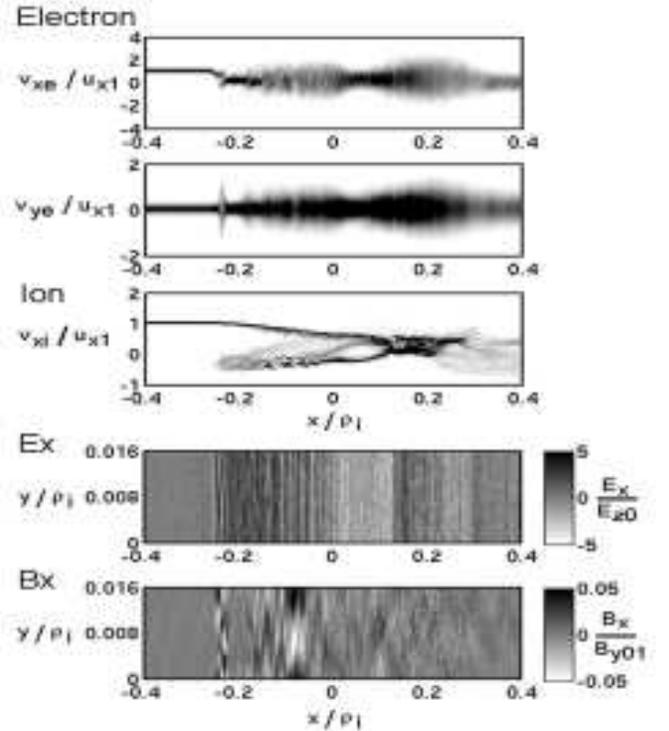}
\caption{Position-velocity phase-space 
distribution functions of electrons and ions in 
$x-v_{xe}$, $x-v_{ye}$, and $x-v_{xi}$ phase spaces. 
at $\omega_{ci1}t = 4.91$, 
and the corresponding spatial profiles of 
$E_x$ and $B_x$ components. 
Note that the $x-v_x$ phase-space distribution functions are reduced 
by integrating over $y$, $v_y$, and $v_z$, 
and the $x-v_y$ phase-space distribution function is reduced 
by integrating over $y$, $v_x$, and $v_z$. 
}
\end{figure}

Figure 3 shows snapshots of position-velocity phase-space 
distribution functions of electrons and ions 
at $\omega_{ci1}t = 4.91$ 
and the corresponding spatial profiles of 
$E_x$ and $B_x$ components. 
At this time, there is a strong ion beam reflected upstream. 
Note that the $x-v_x$ phase-space distribution functions are reduced 
by integrating over $y$, $v_y$, and $v_z$, 
and the $x-v_y$ phase-space distribution function is reduced 
by integrating over $y$, $v_x$, and $v_z$.

Strong heating of electrons
 is found  from the phase-space plots in Fig.~3 
 in the shock-normal direction 
in the shock foot region ($x/\rho_i = -0.2$--0) 
and at the shock overshoot ($x/\rho_i = 0.1$--0.3). 
The shock-normal electric field $E_x$ is strongly enhanced 
by the current-driven instability due to reflected ions 
in the shock foot region. 
An electron phase-space vortex is also identified 
at the leading edge of reflected ions 
($x/\rho_i \sim -0.2$). 
These results are consistent 
with the previous 1D simulations. 

The beam-plasma interaction in the shock-normal direction 
results in strong temperature anisotropy 
$T_{e\perp}/T_{e||}>10$. 
We found enhancement of the shock-normal magnetic component $B_x$ 
especially in the electron phase-space vortex and 
in the shock foot region, 
which corresponds to electromagnetic whistler mode waves 
due to electron temperature anisotropy. 
However, the amplitude of the whistler mode is not so strong 
that the structure of the transverse magnetic field $B_y$ 
is not modified, keeping an almost 1D structure. 

In the electron $x-v_y$ phase-space plot, 
the strong heating of electrons 
is found in the electron phase-space vortex, 
in the shock foot region, and at the shock overshoot. 
These electrons are thermalized 
along the ambient magnetic field 
by the whistler mode wave. 
The electron temperature anisotropy, 
ensuing excitation of electromagnetic whistler mode waves, 
and electron heating along the ambient magnetic field 
would be common features in perpendicular shocks. 
However, these were not seen 
in the previous 1D simulation.


\begin{figure}[t]
\includegraphics[width=0.5\textwidth]{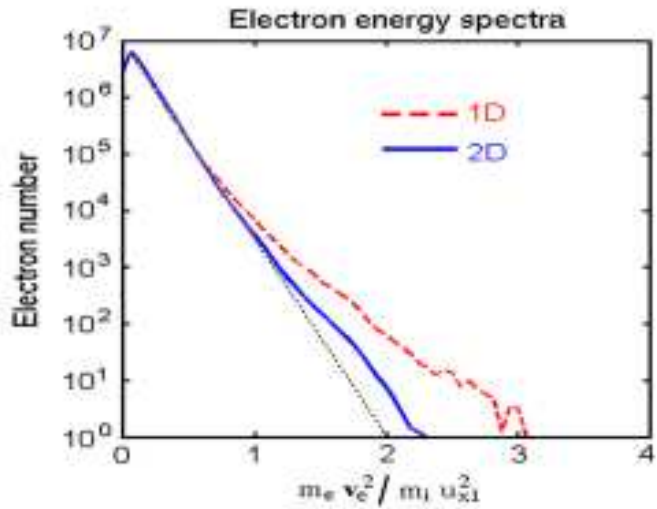}
\caption{
Energy distribution functions of electrons 
in the downstream region 
for 1D and 2D simulations. 
The solid line shows the 2D simulation result 
and the dashed line shows the 1D simulation result. 
The dotted line indicates Maxwellian distribution. 
The energy is normalized by the upstream bulk energy of ions. 
}
\end{figure}

Figure 4 shows energy distribution functions of electrons 
in the downstream region 
for 1D and 2D simulations.
The solid line shows the 2D simulation result,
and the dashed line shows the 1D simulation result. 
It is known that there appear nonthermal electrons 
above the upstream bulk energy of ions 
in 1D simulations 
\citep{Shimada_2000,Hoshino_2002}.
On the other hand, 
we found less nonthermal electrons 
in the present 2D simulation.


The previous 1D studies reported 
electron surfing acceleration with 
coherent electrostatic solitary structures 
in the transition region 
\citep{Shimada_2000,Hoshino_2002,Schmitz_2002a,Schmitz_2002b,Umeda_2006}. 
In the present 2D simulation, 
however, we could not confirm 
the existence of solitary structures in the shock foot region. 
It is well known that 
coherent electrostatic solitary structures 
generated by nonlinear beam-plasma interactions 
need a strong guiding magnetic field 
to maintain an electrostatic equilibrium state, 
the so-called BGK mode \citep{BGK}. 
In the present simulation model, 
electrostatic fields appear in the $x$-direction, 
while a weak ambient magnetic field is taken 
in the $y$-direction. 
Since there is no guiding magnetic field in the $x$-direction, 
electrons can freely move along the ambient magnetic field, 
by which electrostatic waves diffuse.  
As a result, the saturation level of electrostatic waves 
becomes much lower than that with a strong guiding magnetic field
\citep[e.g.,][]{Morse_1969,Muschietti_2000,Umeda_ESW}, 
and the excited electrostatic waves do not develop into 
coherent solitary structures. 

In addition, 
waves can also propagate along the ambient magnetic field 
in the present 2D model. 
Electromagnetic whistler mode waves are 
excited by temperature anisotropy,
which is caused by electron heating in the shock-normal direction 
due to current-driven instability. 
The excited whistler mode waves propagate along the magnetic field 
and thermalize electrons in the direction parallel 
to the ambient magnetic field. 
As a result, 
we found strong heating of electrons along the magnetic field 
instead of electron surfing acceleration across the magnetic field. 
We did not find any further acceleration 
by the whistler mode waves.

\section{Summary and Discussion}

We developed a 2D shock-rest-frame model 
for full particle simulations. 
Then a 2D simulation of a low-beta and 
high Mach number perpendicular shock was performed, 
which shows that 
electrons are thermalized in the direction parallel 
to the ambient magnetic field 
in contrast to the previous 1D simulations. 
The heating of electrons along the ambient magnetic field 
is due to electromagnetic whistler mode waves 
excited by temperature anisotropy. 
In addition, the current-driven instability due to reflected ions 
does not lead to formation of 
coherent electrostatic solitary structures 
because of the free motion of electrons along the ambient 
magnetic field. 
As a result of these consequences, 
the generation of nonthermal electrons via 
electron surfing acceleration due to 
electrostatic solitary structures 
is suppressed in the 2D system. 
This process is independent of shock parameters 
such as $\omega_{pe}/\omega_{ce}$, $\beta$, and $M_A$. 
Even with the realistic parameters, 
there would be less surfing acceleration 
in purely perpendicular shocks, 
because there is no guiding magnetic field 
to stabilize coherent electrostatic structures. 
However, the surfing acceleration might be effective 
in an oblique shock 
in which a guiding magnetic field 
could allow the formation of 
large-amplitude electrostatic structures.

It is noted that 
\citet{Ohira_2007} have also shown 
that electron surfing acceleration becomes absent 
by a simplified 2D simulation with different shock geometry. 
That is, their simulation domain has been taken 
in a plane perpendicular to the ambient magnetic field 
(which corresponds to the $x$-$z$ plane in the present model). 
Thus, electrostatic solitary structures generated 
by the current-driven instability lose coherency 
by a quite different mechanism. 
However, both 2D simulations 
with different axes of coordinate 
have indicated that the 
electron surfing acceleration is suppressed 
more than in 1D simulations. 
Hence, another unknown injection mechanism 
for electron acceleration might be necessary.

\acknowledgments

The computer simulation was carried out 
on Fujitsu HPC2500 at ITC in Nagoya Univ. 
and NEC SX-7 at YITP at Kyoto Univ. 
as a collaborative computational research project
at STEL at Nagoya Univ. and YITP at Kyoto Univ. 
This work was supported in part 
by Grant-in-aid for Creative Scientific Research 
17GS0208 (T.~U.) 
and 
Grant-in-aid from the 
Ministry of Education, Culture, Sports, Science,
and Technology (MEXT) of Japan,
No.~18740153, No.~19047004 (R.~Y.).

\end{document}